\documentclass[aps,prl,reprint,groupedaddress,showpacs]{revtex4-1}

\usepackage{graphicx}
\begin{document}

%\preprint{}

\title{Gigantic Maximum of Nanoscale Noncontact Friction}


\author{Kohta Saitoh}
\author{Kenichi Hayashi}
\author{Yoshiyuki Shibayama}
\author{Keiya Shirahama}
%\thanks{}
\affiliation{Department of Physics, Faculty of Science and Technology, Keio University, Yokohama 223-8522, Japan}


%\date{\today}

\begin{abstract}
We report measurements of noncontact friction between surfaces of NbSe$_{2}$ and SrTiO$_{3}$, and a sharp Pt-Ir tip that is oscillated laterally by a quartz tuning fork cantilever. At 4.2~K, the friction coefficients on both the metallic and insulating materials show a giant maximum at the tip-surface distance of several nanometers. The maximum is strongly correlated with an increase in the spring constant of the cantilever. These features can be understood phenomenologically by a distance-dependent relaxation mechanism with distributed time scales. 
\end{abstract}

%\keywords{}

\maketitle


Friction has been studied for a long time as one of the fundamental subjects in physics.
However, the microscopic mechanism of friction is still in dispute~\cite{persson}. 
Nanotribology, namely study of friction at nanoscale, is the most important subject not only for understanding friction but also for the development of micro- and nanoelectromechanical devices, which need control of friction at the nanoscale. A significant amount of research effort has been devoted to revealing the mechanism of friction at the nanoscale~\cite{nanotribology}.

Interestingly, there is a novel type of friction, so-called noncontact friction, at the nanoscale. 
In contrast to the ordinary contact friction, noncontact friction occurs when two bodies are not in direct contact. It has been observed in scanning probe microscopy experiments, in which a sharp metal tip oscillates laterally near a flat surface~\cite{StoweAPL, StipePRL, KuehnPRL}. Stipe \textit{et al.} observed the noncontact friction between a Au(111) surface and a Au-coated probe tip attached to a very soft cantilever (spring constant $k_{0}\sim 10^{-4}$~N/m)~\cite{StipePRL}. At temperatures 4$<T<$300~K, the friction coefficient $\Gamma$ was approximately $10^{-12}$~kg/s at a tip-sample distance $d<$10~nm. 
As a possible mechanism of the noncontact friction, Ohmic losses caused by fluctuating electromagnetic fields were proposed~\cite{VolokitinPRB1, ChumakPRB}. However, the observed friction coefficient is 7 - 8 orders of magnitude larger than the values derived by the theories. 
Some additional mechanisms that could explain the large noncontact friction have been proposed~\cite{VolokitinRMP, ZuritaPRA}, but the discrepancy has not been solved. 

The above-mentioned theories predict that noncontact friction is proportional to the electrical resistivity of samples. It is therefore expected to be higher on insulating materials than on metals~\cite{VolokitinRMP, ZuritaPRA}.
It was found experimentally that the noncontact friction coefficients of insulating silica and polymer films were an order of magnitude larger than the value on the Au(111) surface~\cite{StipePRL, KuehnPRL}. This tendency is qualitatively consistent with the theories, but quantitative contradiction still remains. On the other hand, Karrai and Tiemann (KT) observed a huge $\Gamma $, which is estimated to be $\sim 10^{-4}$~kg/s, between a conductive graphite surface and a gold probe tip attached to a hard ($k_{0}\sim 10^{4}$~N/m) quartz tuning fork (QTF) at $d<$10~nm at room temperature~\cite{KarraiPRB}. They attributed the origin of the friction to the viscous damping caused by residual adsorbates such as carbon oxide. The friction observed by KT was therefore not discussed in terms of noncontact friction. 
However, it seems implausible that the viscous adsorbates always dominate the friction under high-vacuum conditions. 
It is desirable to perform the measurement of noncontact friction in a more systematic and controlled way with a variety of materials.
It is also worthwhile to study the noncontact friction on superconducting surfaces. Theories predict elimination of noncontact friction on superconductors~\cite{ZuritaPRA}. The friction on superconductors has been also a controversial issue in the past decade~\cite{DayoPRL, PiernoPRL}.

In this Letter, we report the measurements of the lateral friction force between a metal tip and metallic (superconducting) as well as insulating materials that provide clean and flat surfaces. 
We performed the measurements at low temperatures down to 4.2 K under high vacuum ($\sim 10^{-4}$~Pa), by using a lateral force microscope with a hard ($k_{0}\simeq 2.1\times 10^{4}~\mathrm{N/m}$) QTF as a cantilever. The employment of the hard cantilever at low temperatures led us to the discovery of a quite unexpected feature of noncontact friction.

\begin{figure}[t]
  \begin{center}
    \includegraphics[keepaspectratio=true,height=50mm]{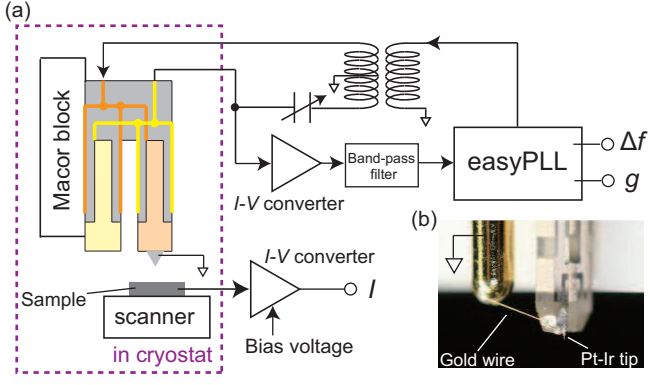}
  \end{center}
  \caption{(a) Experimental setup. The QTF with the Pt-Ir tip oscillates parallel to the sample surface at its resonance frequency with constant amplitude under the control of a commercial FM-AFM controller EasyPLL. The frequency shift $\Delta f$, the dissipation $g$ and the tunneling current $I$ are measured as a function of the tip-sample distance. (b) A photograph of the QTF. The tip is electrically isolated from the QTF and is connected to the ground port by a gold wire.}
  \label{Fig:1}
\end{figure}

We modified a home-made frequency-modulation atomic force microscope (FM-AFM) working at low temperatures~\cite{Saitoh1, Saitoh2} to detect lateral forces. A schematic diagram along with a photograph of the QTF is shown in Fig.~\ref{Fig:1}. Electrochemically etched Pt-Ir tips were attached to one prong of the QTF. The curvature radius of the tip was approximately 100~nm, which is smaller than the value of the soft-cantilever tip used in Ref.~\cite{StipePRL}($\sim 1~\mathrm{\mu m}$). The QTF cantilever is mounted perpendicular to the sample, so the tip oscillates parallel to the sample's surface. The QTF was driven to oscillate at its resonance frequency with constant amplitude by a commercial FM-AFM controller. The resonance frequency of the QTF, $f_{0}$, was approximately 31.6~kHz. In high vacuum ($10^{-4}-10^{-3}$~Pa), the quality factor $Q$ was $6\times 10^{3}$ and $4\times 10^{4}$ at room temperature and 4.2~K, respectively.
 
Throughout our experiments, the friction coefficient $\Gamma _{\mathrm{int}}$ and the friction-induced spring constant $k_{\mathrm{int}}$ defined by the tip-sample interaction were measured as a function of tip-sample distance $d$. The dissipation $g$, which is the output of the automatic gain control circuit in the FM-AFM controller, and the resonance frequency shift $\Delta f$ were collected simultaneously, and converted to  $\Gamma _{\mathrm{int}}$ and $k_{\mathrm{int}}$, respectively: $\Gamma _{\mathrm{int}}= \Gamma _{0} (g/g_{0} - 1)$ and $k_{\mathrm{int}} = 2k_{0}\Delta f /f_{0}$, where $\Gamma _{0} = k_{0}/(2\pi f_{0}Q)$ and $g_{0}$ is the dissipation without the tip-sample interaction. We employed 2$H$-NbSe$_{2}$ and SrTiO$_{3}$ as superconducting ($T_{\mathrm{c}}\simeq $7.2~K) and insulating samples, respectively. NbSe$_{2}$ was cleaved in atmosphere just before being assembled on the microscope. The surface of SrTiO$_{3}$ had been chemically etched by the manufacturer, and we utilized it as purchased. In order to minimize the contaminations on the tip and the samples, the inside of the vacuum can of a cryostat was pumped just after the QTF and the samples were installed in the microscope. The tip-sample distance $d$ was determined by the tunneling current $I$.

\begin{figure}[b]
  \begin{center}
    \includegraphics[keepaspectratio=true,height=95mm]{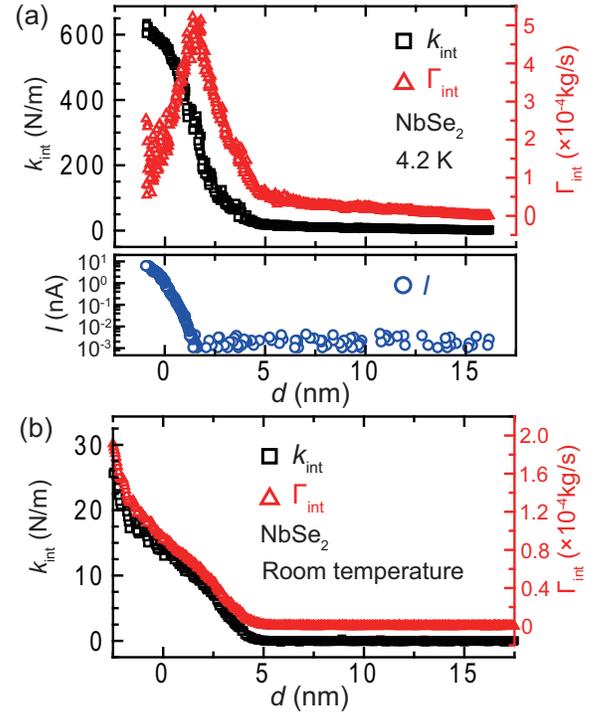}
  \end{center}
  \caption{The noncontact friction coefficient $\Gamma _{\mathrm{int}}$ and the friction-induced spring constant $k_{\mathrm{int}}$ of NbSe$_{2}$ as a function of the tip-sample distance $d$. (a) Data at 4.2~K and under $\sim 10^{-4}$~Pa. Lower panel: the tunneling current $I$. (b) Data at room temperature and under $\sim 10^{-3}$~Pa. 
 %As $d$ decreases, both $\Gamma _{\mathrm{int}}$ and $k_{\mathrm{int}}$ increase monotonically. However, at 4.2~K and $d\simeq $2~nm, $\Gamma _{\mathrm{int}}$ exhibits a maximum. 
 Note the different scales for $\Gamma _{\mathrm{int}}$ and $k_{\mathrm{int}}$ in the two graphs.}
  \label{Fig:2}
\end{figure}

Figure~\ref{Fig:2} shows the $\Gamma _{\mathrm{int}}$ and $k_{\mathrm{int}}$ of NbSe$_{2}$ as a function of $d$, together with $I$. The origin of the tip-sample distance was defined as the distance at which $I$ = 1~nA when the bias voltage applied is equal to 50~mV, i.e. the tunneling resistance is equal to 50~M$\Omega $. This definition was based on prior knowledge that a tunneling resistance of 50~M$\Omega $ is much higher than the quantum resistance $h/2e^{2}$ at which a point contact between the tip and the sample is formed~\cite{RubioPRL}. The electrical conduction at 50~M$\Omega $ was largely dominated by electron tunneling; i.e., the tip was not in physical contact with the sample surface. Several novel features are identified in the data. At 4.2~K, as $d$ decreases, both $\Gamma _{\mathrm{int}}$ and $k_{\mathrm{int}}$ increase for $d<$15~nm. The order of magnitude of the $\Gamma _{\mathrm{int}}$ is $10^{-4}$~kg/s, which is as huge as the value observed by KT~\cite{KarraiPRB}. Most surprisingly, $\Gamma _{\mathrm{int}}$ exhibits a maximum at $d=$1.5~nm, while $k_{\mathrm{int}}$ increases in the same $d$ range. It is notable that at 4.2~K, $\Gamma _{\mathrm{int}}$ is observed on a superconducting state. At $d\simeq $0, $\Gamma _{\mathrm{int}}$ stops to decrease, while $k_{\mathrm{int}}$ tends to remain constant. At room temperature (Fig.~\ref{Fig:2}(b)), $\Gamma _{\mathrm{int}}$ and $k_{\mathrm{int}}$ increase monotonically with decreasing $d$, but $\Gamma _{\mathrm{int}}$ does not show any maximum down to $d=$0.

\begin{figure}[t]
  \begin{center}
    \includegraphics[keepaspectratio=true,height=80mm]{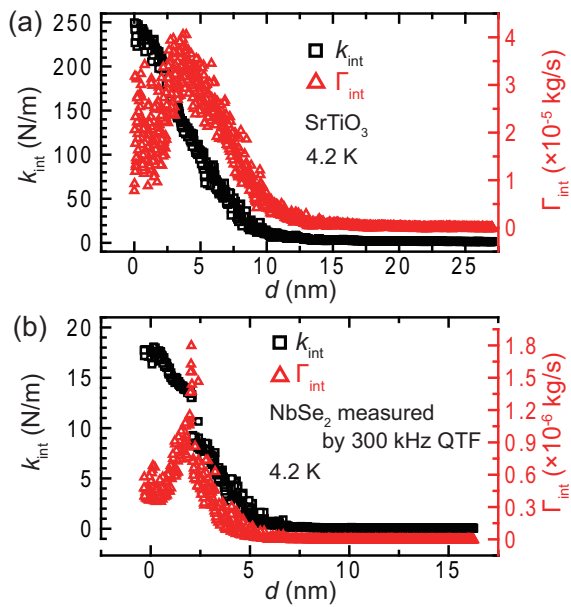}
  \end{center}
  \caption{(a) $\Gamma _{\mathrm{int}}$ and $k_{\mathrm{int}}$ as a function of $d$ for the SrTiO$_{3}$ sample at 4.2~K. Note that both $\Gamma _{\mathrm{int}}$ and $k_{\mathrm{int}}$ vary over a wider distance range (about 25~nm) than those of NbSe$_{2}$. (b) $\Gamma _{\mathrm{int}}$ and $k_{\mathrm{int}}$ for NbSe$_{2}$ at 4.2~K measured by a QTF oscillating at 300~kHz. $\Gamma _{\mathrm{int}}$ and $k_{\mathrm{int}}$ are 2 and 1 order of magnitude smaller, respectively, than those taken at 32~kHz.}
  \label{Fig:3}
\end{figure}

Similar patterns are observed in the data collected with SrTiO$_{3}$. Figure~\ref{Fig:3}(a) shows $\Gamma _{\mathrm{int}}(d)$ and $k_{\mathrm{int}}(d)$ at 4.2~K. Since SrTiO$_{3}$ is an insulator, the smallest tip-sample distance at which reasonable data are available has been determined to be the zero of $d$. The overall behavior, which is a maximum of $\Gamma _{\mathrm{int}}$ associated with an increase in $k_{\mathrm{int}}$, is quite similar to the results obtained with NbSe$_{2}$. We found that both $\Gamma _{\mathrm{int}}$ and $k_{\mathrm{int}}$ start to increase at a relatively longer distance of $d\sim $25~nm than in the case of NbSe$_{2}$. The magnitude of $\Gamma _{\mathrm{int}}$ is approximately $10^{-5}$~kg/s. This value is an order of magnitude smaller than the $\Gamma _{\mathrm{int}}$ observed in NbSe$_{2}$.
 
Figure~\ref{Fig:3}(b) shows the data collected by using NbSe$_{2}$ at 4.2~K and a different QTF, oscillating at a resonance frequency of 300~kHz. The definition of the zero of $d$ is the same as in Fig.~\ref{Fig:2}. Once again, $\Gamma _{\mathrm{int}}$ exhibits a maximum associated with an increase in $k_{\mathrm{int}}$ at $d<$10~nm. We also found that $k_{\mathrm{int}}$ changes abruptly at $d\sim $2~nm. This abrupt change was not due to an irreversible change of the tip, e.g. plastic deformation or the damage in the tip, because the abrupt change always occurred at a single tip-sample distance, and it was reproduced at different lateral positions. $\Gamma _{\mathrm{int}}$ is approximately $10^{-6}$~kg/s, which is 2 orders magnitude smaller than the value obtained with the QTF oscillating at 32~kHz. On the other hand, $k_{\mathrm{int}}$ is about 10~N/m, namely, an order of magnitude smaller than the $k_{\mathrm{int}}$ obtained with the 32~kHz QTF.

The maxima of $\Gamma _{\mathrm{int}}(d)$ and the increase in $k_{\mathrm{int}}(d)$ were reproducibly observed in superconducting and insulating materials over a wide range of frequencies. Furthermore, on NbSe$_{2}$, the behaviors were also observed at temperatures above the superconducting transition temperature 7.2~K up to 10~K. These facts definitely show that the $\Gamma _{\mathrm{int}}$ maximum associated with the $k_{\mathrm{int}}$ increase is an universal feature of noncontact friction.
We can also conclude that the conductivity of the materials has negligible contribution to the observed giant noncontact friction. The mechanism underlying the behavior of our $\Gamma _{\mathrm{int}}$ is expected to be different from the existing theoretical notions providing an interpretation of noncontact friction.

We will argue below that the $\Gamma _{\mathrm{int}}$ maximum with the $k_{\mathrm{int}}$ increase is an intrinsic property of noncontact friction. First, in the NbSe$_{2}$ data, the $\Gamma _{\mathrm{int}}$ maximum was observed at $d\sim $2~nm, where no tunneling current was detected. Therefore, the possibility that the $\Gamma _{\mathrm{int}}$ maximum was caused by the tip coming in physical contact with the sample is excluded. Second, during the measurements the oscillation amplitude of the QTF was kept constant to about 0.3~nm, which is close to the lattice constant of NbSe$_{2}$. It is much smaller than $d\sim $15~nm, where $\Gamma _{\mathrm{int}}$ starts to increase. Moreover, the behaviors of $\Gamma _{\mathrm{int}}$ and $k_{\mathrm{int}}$ were reproduced for many different lateral positions on the samples' surface. Therefore, we conclude that there was no intermittent contact in the lateral direction, i.e., no instantaneous collisions of the probe tip to protrusions or steps on the samples' surface. The possibility of physical contact of the tip with the surfaces is excluded once again. Third, the possibility of energy dissipation due to viscous adsorbates that KT claimed~\cite{KarraiPRB} is excluded, because all adsorbates, if there are such materials present, freeze at low temperatures. In addition, helium atoms, which exceptionally remain in liquid form, did not exist in the experimental apparatus, because the microscope was cooled down to 4.2~K without using any thermal exchange gas such as helium. On the grounds of these observations, we conclude that the anomalies identified in $\Gamma _{\mathrm{int}}(d)$ and $k_{\mathrm{int}}(d)$ reflect a novel, universal feature of nanoscale noncontact friction.
 
KT observed a minimum in $Q$, which corresponded to a maximum in $\Gamma _{\mathrm{int}}$, only when they used a long probe tip and hence the spring constant of the tip was 2 orders of magnitude smaller than that of their QTF~\cite{KarraiPRB}. In this case, the minimum in $Q$ is attributed to an experimental artifact caused by the deformation of the soft tip. In our experiment, the spring constant of the probe tip is an order of magnitude larger than that of the QTF. Therefore, our $\Gamma _{\mathrm{int}}$ maximum is not an artifact but an intrinsic property of noncontact friction.

\begin{figure}[t]
  \begin{center}
    \includegraphics[keepaspectratio=true,height=60mm]{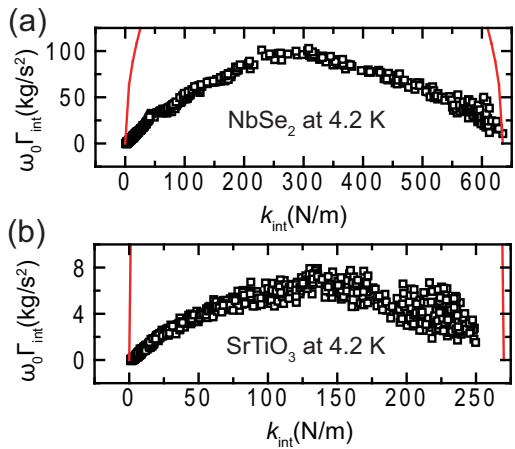}
  \end{center}
  \caption{The $k_{\mathrm{int}} - \omega_0 \Gamma _{\mathrm{int}}$ plots: (a) NbSe$_2$ data of Fig.~\ref{Fig:2}(a). (b) SrTiO$_3$ data from Fig.~\ref{Fig:3}(a). Note that the ellipse of the SrTiO$_{3}$ data is more squeezed in the vertical direction by an order of magnitude than that for NbSe$_{2}$. Red lines are a part of arbitral semicircles.}
  \label{Fig:4}
\end{figure}

We noticed that the overall behavior of $\Gamma _{\mathrm{int}}$ and $k_{\mathrm{int}}$ bear a striking resemblance to the Debye-like relaxation response function of dielectrics~\cite{dielectrics}. It is therefore reasonable to explain our noncontact friction in terms of the characteristic timescale of some relaxation mechanism taking place on the samples' surface. More specifically, the behavior of $\Gamma _{\mathrm{int}}$ as well as of $k_{\mathrm{int}}$ can be simultaneously determined based on the relaxation time $\tau $ that depends on the tip-sample distance $d$.
Based on this idea, we replotted the data shown in Fig.~\ref{Fig:2}(a) and Fig.~\ref{Fig:3}(a) on a $k_{\mathrm{int}} - \omega _{0}\Gamma _{\mathrm{int}}$ plane, as shown in Fig.~\ref{Fig:4}. This is similar to the so-called Cole-Cole plot for dielectric phenomena~\cite{Cole-Cole}, because $\Gamma _{\mathrm{int}}$ and $k_{\mathrm{int}}$ are represented as $\Gamma _{\mathrm{int}}\dot{x}+k_{\mathrm{int}}x=(k_{\mathrm{int}}+i\omega \Gamma _{\mathrm{int}})x\cong (k_{\mathrm{int}}+i\omega _{0}\Gamma _{\mathrm{int}})x\equiv G(\omega _{0})x$ in the equation of motion of a cantilever, where $x$ is the displacement of the cantilever.
Here $G(\omega _{0})$ corresponds to the response function. The $k_{\mathrm{int}} - \omega _{0}\Gamma _{\mathrm{int}}$ plots for NbSe$_{2}$ and SrTiO$_{3}$ exhibit an ellipse rather than a semicircle. When we compared these plots with the results of a Cole-Cole analysis~\cite{Cole-Cole}, it was made evident that the ellipses indicate that the relaxation process cannot be described by a single time scale but by a wide distribution of relaxation times. It is worth noting that the ellipse for NbSe$_{2}$ is wider than that for SrTiO$_{3}$. This observation led us to the conclusion that the relaxation time depends on the materials and is more distributed for SrTiO$_{3}$ than for NbSe$_{2}$. In addition, the $d$ dependence of the relaxation time changes with temperature. The absence of a maximum for $\Gamma _{\mathrm{int}}$ in the case of NbSe$_{2}$ at room temperature can be attributed to the change of the $d$ dependence with temperature and the possibility that the probe tip comes in contact with the sample surface before $\Gamma _{\mathrm{int}}$ reaches its maximum.

The next step is to propose a model to describe the relaxation mechanism. The lateral oscillation of the probe tip causes the crystal lattice of the sample surface to deform and the surface atoms to vibrate. It is natural to assume that the sample atoms are vibrated more easily as $d$ decreases, so the vibration of the sample atoms following the tip depends on $d$. Given that the $d$-dependent sliding friction forces drive the vibratory motion of the atoms on the samples' surface, we propose to model the lateral force interaction by a dashpot $\gamma _{\mathrm{f}}$ connected in series with a spring $k_{\mathrm{f}}$. In other words, $k_{\mathrm{int}}$ and $\Gamma _{\mathrm{int}}$ are represented as follows: $k_{\mathrm{int}}=k_{\mathrm{f}}\omega _{0}^{2}\tau ^{2}/(1+\omega _{0}^{2}\tau ^{2})$ and $\omega _{0}\Gamma _{\mathrm{int}}=k_{\mathrm{f}}\omega _{0}\tau /(1+\omega _{0}^{2}\tau ^{2})$,
where $\tau $ is the relaxation time and is defined as $\tau =\gamma _{\mathrm{f}}/k_{\mathrm{f}}$. Assuming that $\tau $ increases as $d$ decreases, the behaviors of $\Gamma _{\mathrm{int}}$ and $k_{\mathrm{int}}$ in our experiments can be explained consistently. In reality, the lattice deformation depends on the lateral position of the atoms on the samples' surface and furthermore, the apex of the tip might be deformed. Therefore, the lattice deformation should be described by a distribution of relaxation times. The deviation from the semicircle in the Cole-Cole plot can originate from such a distribution.

The above-mentioned lattice deformation might be associated with the so-called phonon friction, which was proposed as a potential origin of noncontact friction~\cite{VolokitinRMP}. However, the order of magnitude of the friction due to the lattice deformation mediated by van der Waals or electrostatic forces is calculated to be $10^{-18}-10^{-16}$~kg/s~\cite{VolokitinRMP}. Clearly, the giant noncontact friction observed during our experiments cannot be explained by the phonon friction and other theories proposed so far. New theoretical concepts are required to understand the microscopic origin of the noncontact friction.

Finally, we comment on the effect of superconductivity on the noncontact friction. We found no change in both $\Gamma _{\mathrm{int}}$ and $k_{\mathrm{int}}$ at the $T_{\mathrm{c}}$ of NbSe$_2$ ($\simeq $7.2~K) within the experimental accuracy. This shows again that the conducting electrons have a negligible contribution to the giant noncontact friction. However, this does not deny the possibility of the contribution of superconductivity to noncontact friction: It might be masked by the giant friction. Experiments under a magnetic field may reveal the superconducting effect by comparing the friction inside and outside the vortex cores.

In summary, we discovered that the noncontact friction probed by a hard quartz tuning fork cantilever shows a giant maximum at low temperatures. The friction maximum is associated with a change in the friction-induced spring constant of the cantilever. This associated behavior is phenomenologically described by a Debye-like relaxation mechanism with multiple time scales.
Our results show that there are some hidden mechanisms producing the gigantic maximum of the noncontact friction. The abrupt change in the friction-induced spring constant that was observed in the 300~kHz QTF might be a clue for elucidating the mechanisms of the giant noncontact friction. 
Studies of the giant noncontact friction will contribute to understanding general friction phenomena and to the development of nanotribology.

\begin{acknowledgments}
The authors thank Koki Takita for the synthesis of NbSe$_{2}$. This work was supported by a Grand-in-Aid for Scientific Research from MEXT, Japan.
\end{acknowledgments}

%\bibliography{Saitoh_manuscript}
%

\end{document}